\documentclass[aps,showpacs,twocolumn,prb]{revtex4-1}
\usepackage{hyperref}
\usepackage{graphicx}
\usepackage{amsmath,amssymb,amsfonts}

\begin{document}

\title{Anisotropic magnetic properties of the ferromagnetic semiconductor CrSbSe$_3$}
\author{Tai Kong}  
\author{Karoline Stolze}  
\author{Danrui Ni}
\author{Satya K. Kushwaha}
\author{Robert J. Cava}
\affiliation{Department of Chemistry, Princeton University, Princeton, NJ 08544, USA}

\begin{abstract}

Single crystals of CrSbSe$_3$, a structurally pseudo-one-dimensional ferromagnetic semiconductor, were grown using a high-temperature solution growth technique and were characterized by x-ray diffraction, anisotropic, temperature- and field-dependent magnetization, temperature-dependent resistivity and optical absorption measurements. A band gap of 0.7 eV was determined from both resistivity and optical measurements. At high temperatures, CrSbSe$_3$ is paramagnetic and isotropic with a Curie-Weiss temperature of $\sim$145 K and an effective moment of $\sim$4.1 $\mu_B$/Cr. A ferromagnetic transition occurs at $T_c$ = 71 K. The $a$-axis, perpendicular to the chains in the structure, is the magnetic easy axis, while the chain axis direction, along $b$, is the hard axis. Magnetic isotherms measured around $T_c$ do not follow the behavior predicted by simple mean field critical exponents for a second order phase transition. A tentative set of critical exponents is estimated based on a modified Arrott plot analysis, giving $\beta\sim$0.25, $\gamma\sim$1.38 and $\delta\sim$6.6.

\end{abstract}
\maketitle

\section{Introduction}

Magnetic semiconductors have been studied for several decades for their potential applications in spintronics\cite{Sarma04}. Bulk materials like europium and chromium chalcogenides\cite{Dietl02} are early examples of a ferromagnetic-semiconducting state, while ferromagnetism has also been successfully induced in well-developed semiconductors by magnetic doping\cite{Furdyna88,Ohno98}. Another approach in spintronics is to deposit thin layers of a semiconducting ferromagnetic material on top of a non-magnetic material. Recently, there have been many attempts to control magnetism in 2-dimensional (2D) materials or topological insulators by proximitizing a ferromagnetic semiconductor\cite{Wang15PRL,Ji13}. Several 2D magnetic semiconductors, CrGeTe$_3$ and CrI$_3$ in particular, have even been shown to maintain their ferromagnetism at an atomic-layer level\cite{Lin16,Gong17,Huang17b}.

In the Cr$MX_3$ ternary chromium tri-chalcogenides (where $M$ is a non-transition-metal and $X$ = S, Se, Te), the crystal structure varies depending on how Cr$X_6$ octahedra are arranged. Cr$M$Te$_3$ for $M$ = Si, Ge and Sn, for example, have layered structure, with CrTe$_6$ octahedra forming a honeycomb lattice. When $M$ = Sb, Ga, on the other hand, Cr$MX_3$ compounds exhibit a pseudo-one-dimensional crystal structure, with Cr$X_6$ octahedra forming infinite, edge-sharing, double rutile chains, and $M$ atoms linking neighboring chains; for different $M$ atoms, the relative angle between double rutile chains changes\cite{Volkov97}. Among these pseudo-one-dimensional compounds, CrSbSe$_3$ is of interest due to its semiconducting ferromagnetic ground state. Previously, structural and magnetic properties have been reported for polycrystalline samples\cite{Odink93,Volkov97}. Those studies showed that Cr in CrSbSe$_3$ appears as high-spin Cr$^{3+}$, with $S$ = 3/2 and that the material becomes ferromagnetic below $\sim$ 70 K. To better understand the ferromagnetic properties of CrSbSe$_3$ and inspect the influence of structural low dimensionality on its magnetism, we present here electric transport and anisotropic magnetic properties of single crystalline CrSbSe$_3$.

\section{Experimental Methods}

\begin{figure}[tbh!]
\includegraphics[scale = 0.45]{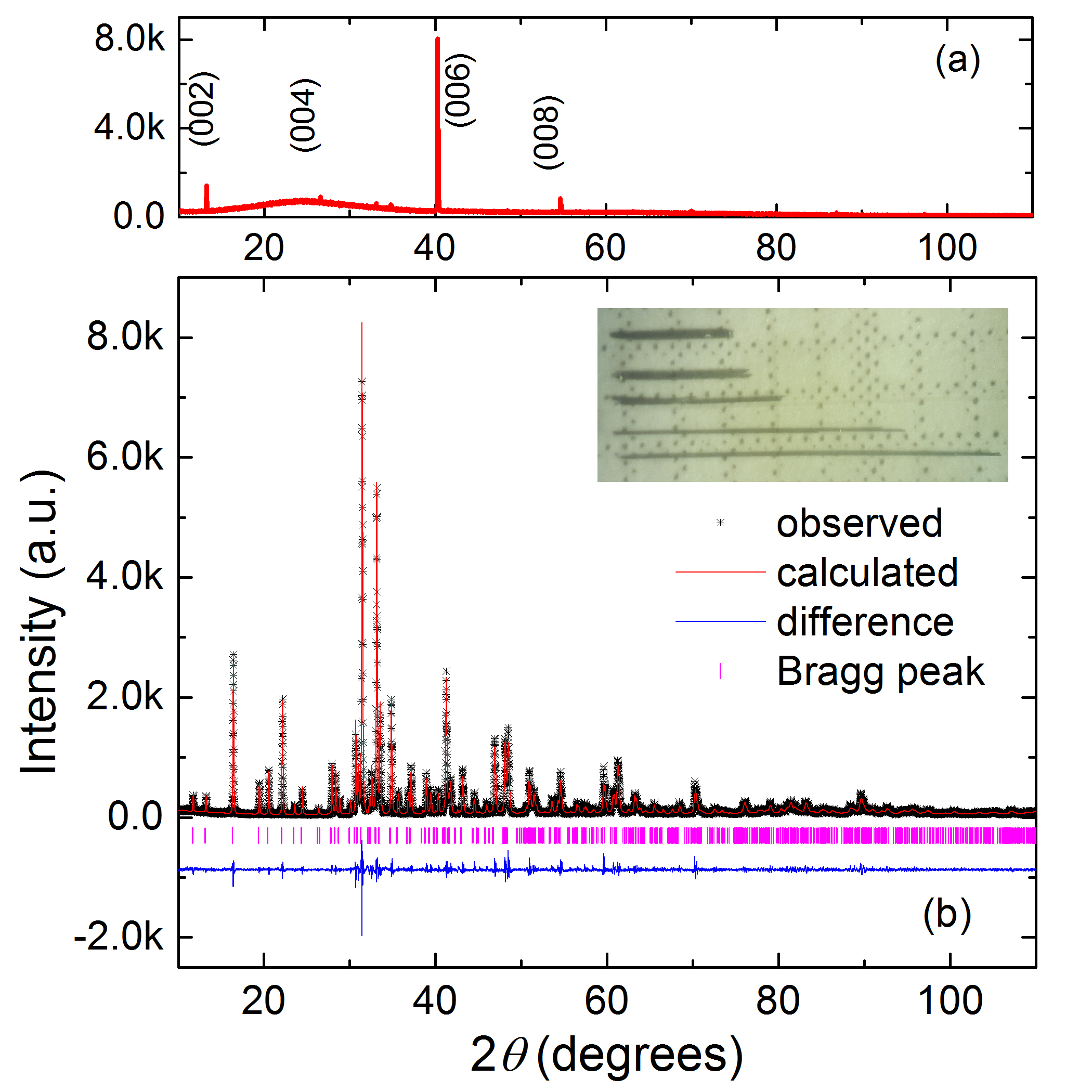}
\caption{(Color online) (a) Single crystal x-ray diffraction pattern along (00$l$). (b) Powder x-ray diffraction pattern of CrSbSe$_3$. Black crosses are measured data; red line shows the LeBail fitted curve; blue line indicates the difference between measured and calculated values; magenta ticks indicates the Bragg reflection positions. The inset shows several single crystals of CrSbSe$_3$ on a millimeter grid paper.}
\label{xray}
\end{figure}

CrSbSe$_3$ single crystals were grown by crystallization from a Se-rich solution. Starting bulk elements were mixed in a molar ratio of Cr:Sb:Se = 7:33:60 and were sealed in an evacuated silica tube. The ampoule was then heated to 800 $^{\circ}$C and slowly cooled to 680 $^{\circ}$C, where the molten liquid was separated from the crystals in a centrifuge with silica wool serving as a filter. Remaining flux that attached to the surface of crystals were removed by keeping the crystal at 500 $^{\circ}$C for 3 days in a sealed silica tube while leaving the cold end of the tube at room temperature. Millimeter-long single crystals of CrSbSe$_3$ are blade-like and malleable (see Fig.\ref{xray}(b)). Polycrystalline CrSbSe$_3$ was also synthesized, via solid state reaction. Starting elements in powder form were mixed in a stoichiometric ratio and sealed in an evacuated silica tube and were kept at 600 $^{\circ}$C for 2 days. 

Magnetization data were measured using a Quantum Design (QD) physical property measurement system (PPMS) Dynacool, equipped with a VSM option. Anisotropic magnetization data were obtained on selective pieces of single crystals. For field-dependent magnetization, when $H\parallel a$ and $c$, single crystals were mounted on a silica sample holder with GE varnish. For $H\parallel$b and for polycrystalline sample, the standard QD plastic capsule was used. Because the mass of each single crystal is small, anisotropic magnetization measurements were normalized to the saturation magnetization values obtained on polycrystalline sample (which has a much larger mass) at 1.8 K. Anisotropic, temperature-dependent magnetization were measured on a shaft of samples, which only distinguish between $H\parallel b$ and $H\perp b$. Temperature-dependent resistivity was measured on a polycrystalline CrSbSe$_3$ pellet sample using the QD electrical transport option (ETO). Pt wires were attached to the sample via DuPont 4922N silver paint, using a two-probe configuration, suitable due to the very high resistance of the material.

\begin{figure}[tbh!]
\includegraphics[scale = 0.43]{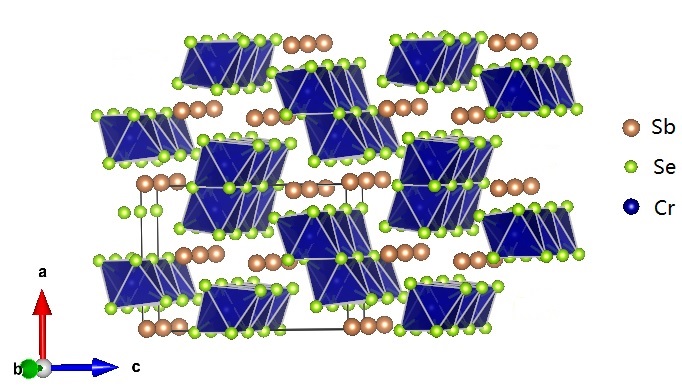}
\caption{(Color online) Crystal structure of CrSbSe$_3$. CrSe$_6$ polyhedra are shown in blue.}
\label{strc}
\end{figure}

Diffuse reflectance spectra were collected by a Cary 6000i UV-VIS-NIR spectrometer with an integrating sphere and were converted from reflectance to absorbance using Kubelka-Munk method. KBr was employed as the substrate. The sample for band gap measurement was mixed with KBr powder (sample to KBr, mass ratio 1:20) and pressed into a pellet. A pure KBr pellet with the same mass was prepared as the blank reference. Powder x-ray diffraction data were collected using a Bruker D8 Advance Eco, Cu K$_{\alpha}$ radiation ($\lambda$ = 1.5406 $\text{\AA}$), equipped with a LynxEye-XE detector. The crystal structure of CrSbSe$_3$ was further determined by single-crystal x-ray diffraction (SXRD), which is presented in the Appendix. 

\section{Results and Discussion}

Powder x-ray diffraction data for CrSbSe$_3$ are shown in Fig.~\ref{xray}(b). All peaks can be indexed by the previously reported crystal structure, in agreement with our SXRD data\cite{Odink93,Volkov97}. For single crystalline samples, the $b$-axis, is along the long crystal dimension\cite{Volkov97}. The $c$-axis direction relative to the crystal morphology was determined from a single crystal laying on its flat face on a glass slide on a powder diffractometer. As shown in Fig.~\ref{xray}(a), these data show predominantly (00$l$) peaks, indicating the $c$-axis is perpendicular to the blade-shaped sample. Minor peaks from other orientations may due to an imperfect alignment of the crystal on the glass slide.

The crystal structure of CrSbSe$_3$ is shown in Fig.~\ref{strc}. Edge sharing, slightly distorted, CrSe$_6$ octahedra form chains that extend along the $b$-axis. Each chain is composed of two parallel, edge sharing columns of CrSe$_6$ octahedra, forming double rutile chains. Between the chains, the Cr atoms are linked by two Se atoms; both Cr-Se-Cr angles are $\sim$92$^{\circ}$. Sb atoms are located in between the double chains. Crystallographic data obtained from SXRD are summarized in Table~\ref{table1} , final atomic parameters are listed in Table~\ref{table2} and~\ref{table3} in the Appendix. 

\begin{figure}[tbh!]
\includegraphics[scale = 0.34]{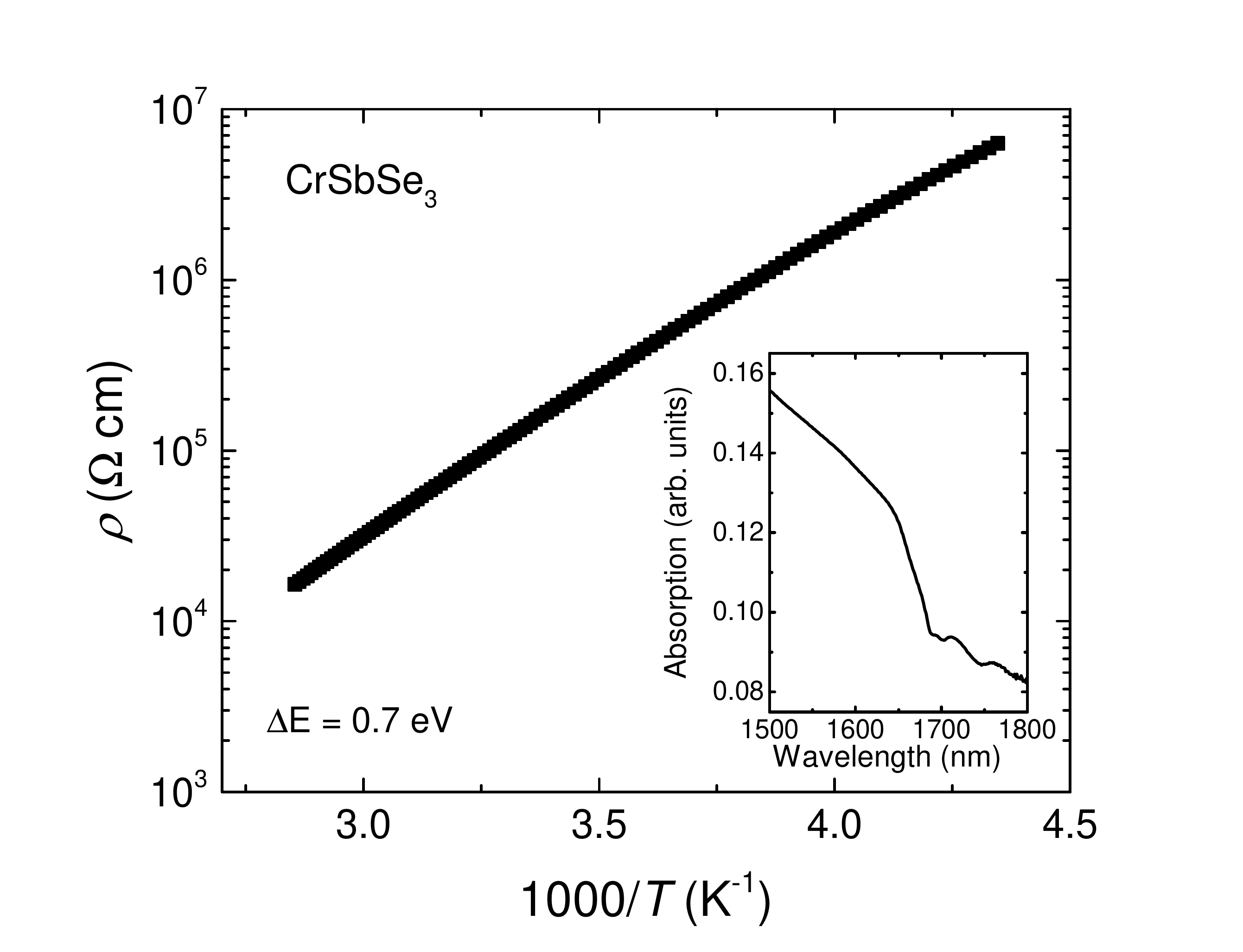}
\caption{Resistivity of CrSbSe$_3$ as a function of inverse temperature on a semi-log plot. Inset shows the optical absorption as a function of wavelength.}
\label{RT}
\end{figure}

The temperature-dependent resistivity of CrSbSe$_3$ is shown in Fig.~\ref{RT}. At 300 K, the resistivity is around 0.1 M$\Omega$ cm and increases with decreasing temperature. When plotting resistivity as a function of inverse temperature on a semi-log plot, the resistivity shows a linear behavior, consistent with semiconducting behavior and a band gap of $\sim$0.7 eV. This band gap was also confirmed with optical obsorbance measurements. In the inset of Fig.~\ref{RT}, the optical absorption coefficient as a function of wavelength is shown. A clear absorption edge is seen at around 1700 nm. The size of the band gap is estimated from this data to be $\sim$0.6 eV based on an indirect band gap\cite{Jain2013} model\cite{Tarasova12}, in agreement with our resistivity result. The measured band gap is slightly larger than the value calculated by DFT\cite{Jain2013}.

\begin{figure}[tbh!]
\includegraphics[scale = 0.35]{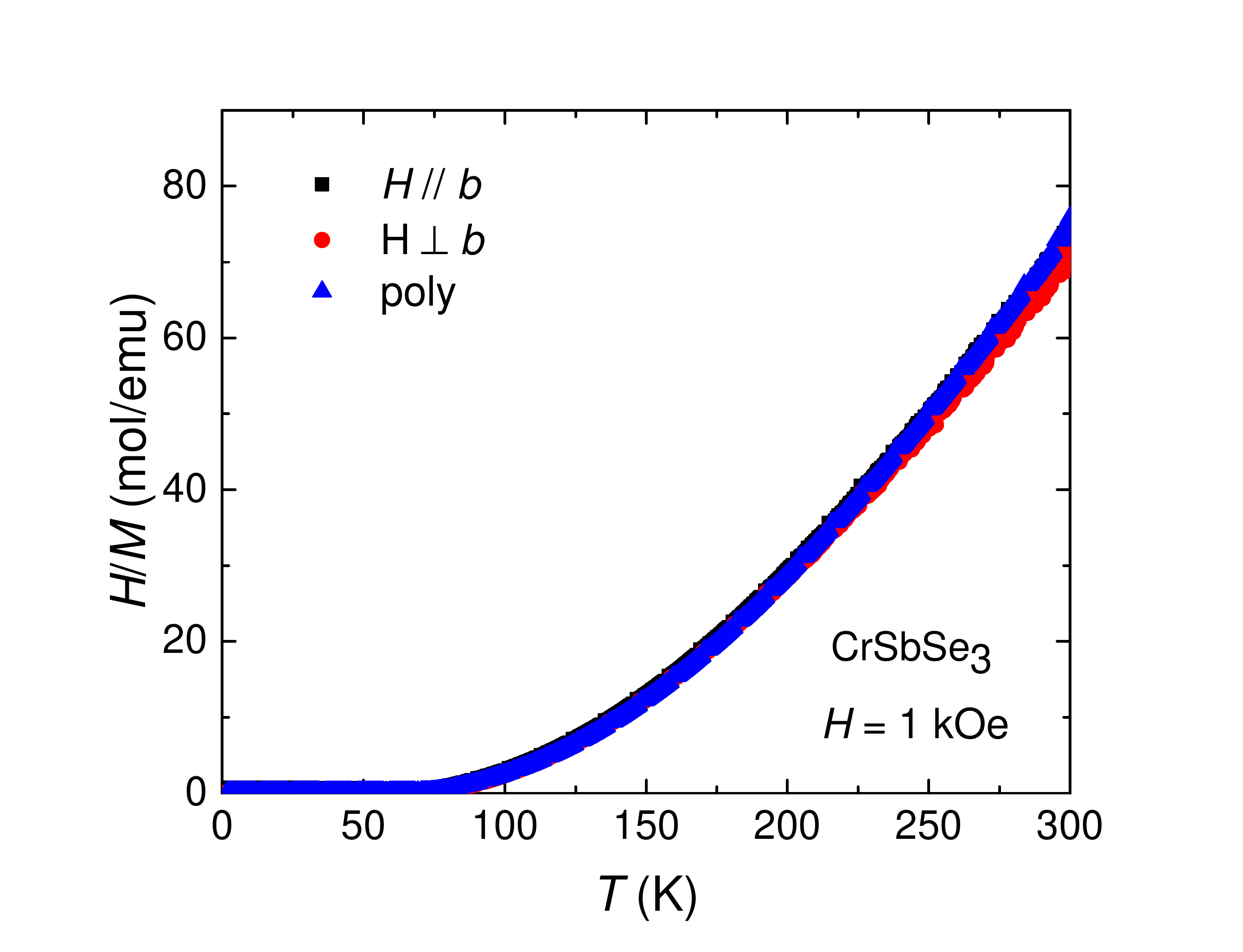}
\caption{(Color online) Anisotropic, temperature-dependent inverse magnetization of CrSbSe$_3$ measured at 1 kOe.}
\label{MT}
\end{figure}

Fig.~\ref{MT} shows the anisotropic magnetization of CrSbSe$_3$ as a function of temperature. In the paramagnetic state, the magnetic susceptibility of CrSeSe$_3$ appears to be nearly isotropic. A linear fit to the high temperature inverse magnetic susceptibility of the polycrystalline data gives an effective moment of 4.1 $\mu_B$/Cr, which is slightly larger than the expected value for Cr$^{3+}$ (3.9 $\mu_B$/Cr). The Curie-Weiss temperature extrapolated from the paramagnetic state is $\sim$145 K. A ferromagnetic transition occurs at around 70 K. Both the effective moment and Curie-Weiss temperature obtained in the current study agree with previously reported data for polycrystalline CrSbSe$_3$ samples\cite{Odink93,Volkov97}.

\begin{figure}[tbh!]
\includegraphics[scale = 0.35]{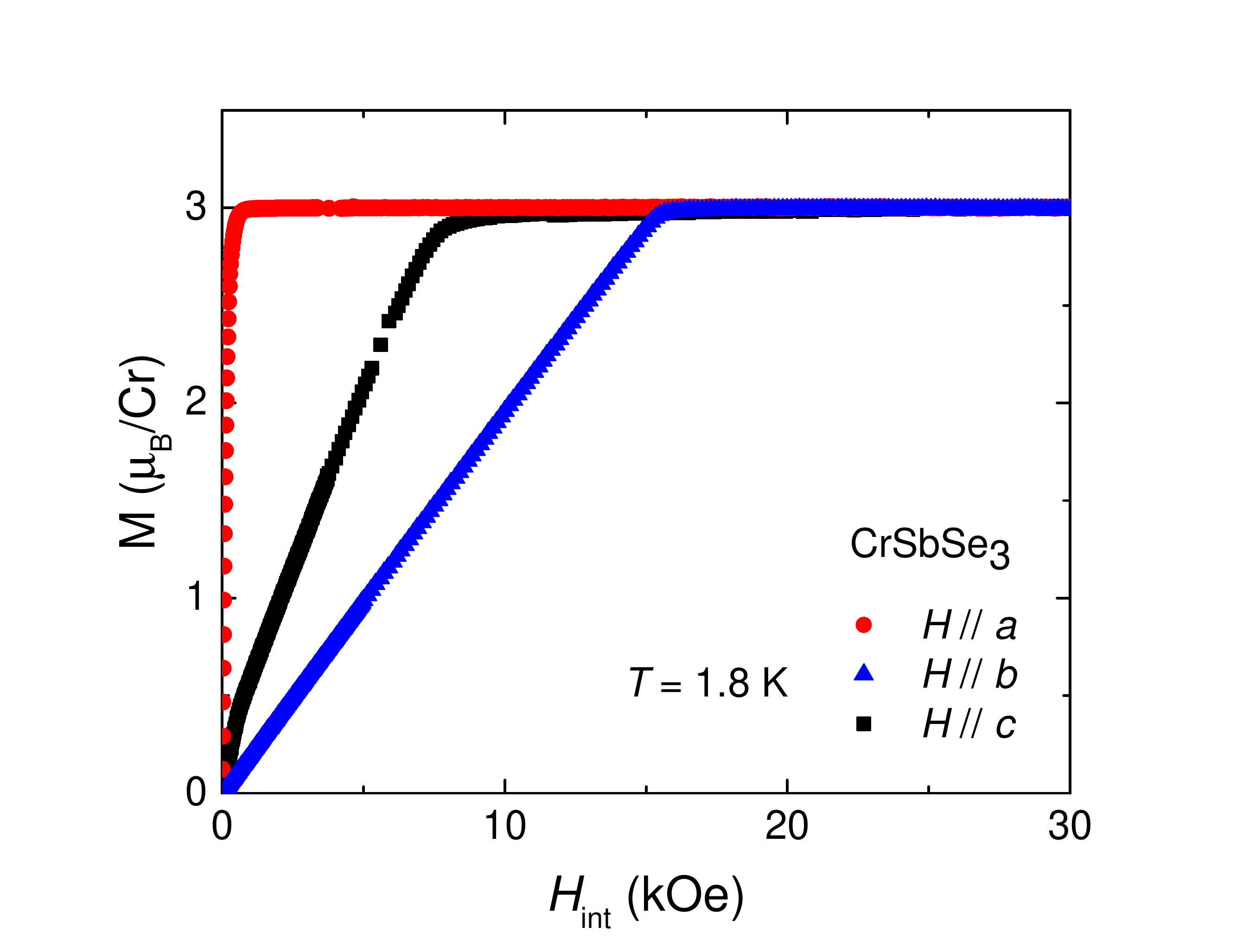}
\caption{(Color online) Anisotropic, field-dependent magnetization of CrSbSe$_3$ measure at 1.8 K. The CrSe$_6$ double rutile chains extend along the $b$-axis.}
\label{MH}
\end{figure}

Fig.~\ref{MH} shows the anisotropic magnetization isotherm measured at 1.8 K. The demagnetization factor was estimated by assuming the geometry of the sample is close to a rectangular shape\cite{Aharoni}. The internal field, $H_{int}$, was calculated from the relation $H_{int} = H - 4\pi NM$, where $N$ is the demagnetization factor. For $H\parallel a$, which is the easy axis, the demagnetization factor was also confirmed with the Arrott plot measured at around 2 K\cite{Arrott, Lamichhane2016}. Both demagnetization factor values are numerically similar. The value obtained from the Arrott plot along the $a$-axis was used to correct the calculated values for the other two orientations. As shown in Fig.~\ref{MH}, the $a$-axis, which is perpendicular to the chains, is the easy axis and the $b$-axis, the direction of the chains, is the hard axis, which saturates at $\sim$15 kOe. The magnetization value saturates at 3 $\mu_B$/Cr, consistent with expectations for $S$ = 3/2 Cr$^{3+}$. 

\begin{figure}[tbh!]
\includegraphics[scale = 0.6]{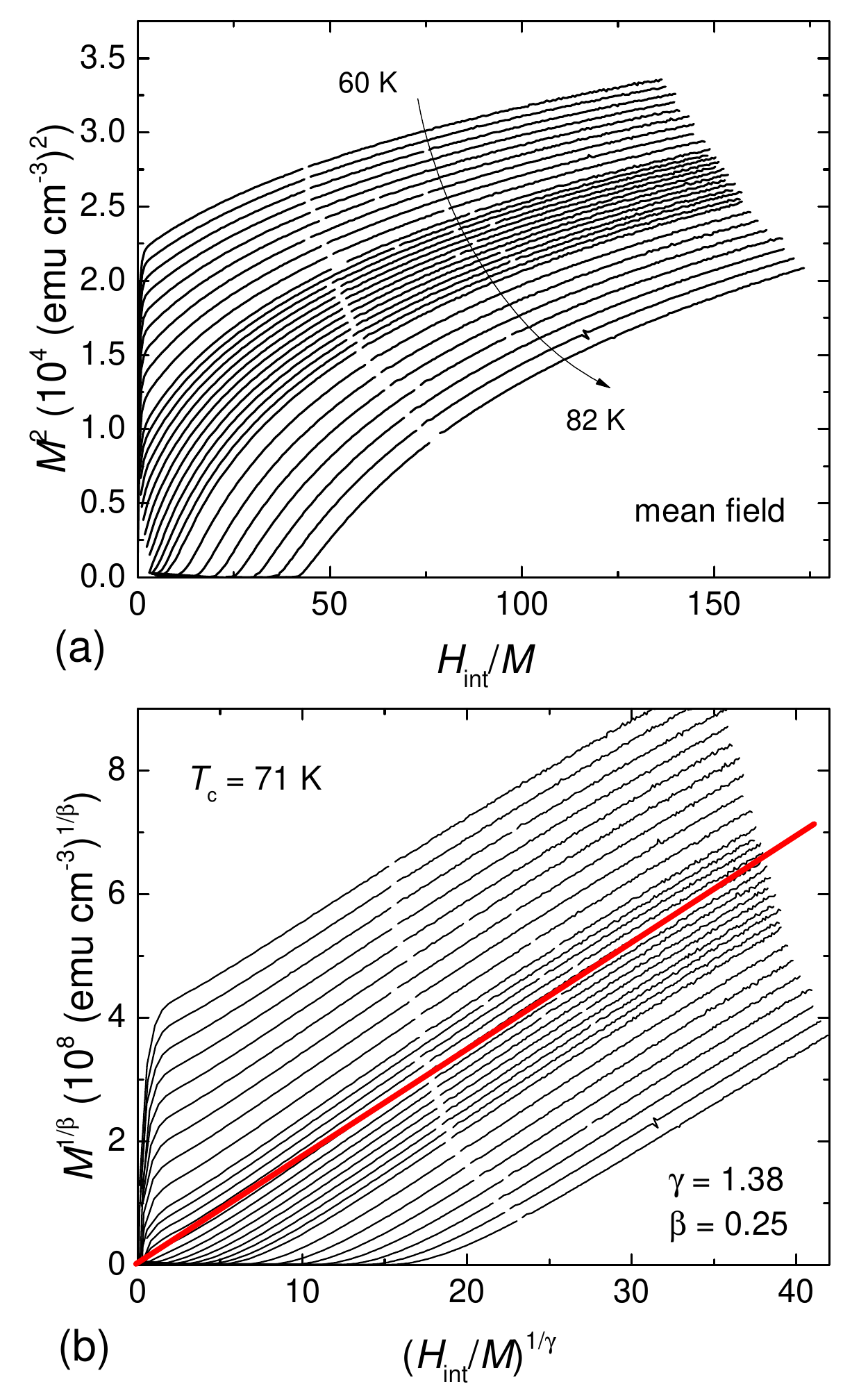}
\caption{(Color online) (a) Arrott plot with critical exponents from mean field theory ($\beta$ = 0.5, $\gamma$ = 1.0). (b) Modified Arrott plot with $\beta$ = 0.25, $\gamma$ = 1.38. Red solid line is a guide for the eyes, indicating the straight line that passes the origin at $T_c$.}
\label{AT1}
\end{figure}

To better determine the ferromagnetic transition temperature, we first considered the well-known Arrott plot\cite{Arrott}. Magnetization isotherms along the easy axis were measured at various temperatures in the vicinity of the ferromagnetic transition temperature. $M^2$ as a function of $H_{int}/M$ is displayed in Fig.~\ref{AT1}(a). Because the Arrott plot describes the magnetic behavior at low fields in the proximity of $T_c$, only magnetization data below 25 kOe were considered. In the mean field description of the magnetization near $T_c$, curves in the Arrott plot should be a set of parallel straight lines with the one passing the origin indicating the ferromagnetic transition temperature. It is clear that the mean field critical exponent does not work for CrSbSe$_3$, as illustrated by a set of curved lines shown in Fig.~\ref{AT1}(a). 

For a second order phase transition, the spontaneous magnetization ($M_s$) below $T_c$, the initial magnetic susceptibility ($\chi^{-1}_0$) above $T_c$ and the field-dependent magnetization ($M$) at $T_c$ are:

\begin{equation}
M_s(T) = M_0(-\epsilon)^{\beta},\ for\ T < T_c,
\end{equation}

\begin{equation}
\chi^{-1}_0 = (h_0/m_0) \epsilon^{\gamma},\ for\ T > T_c,
\end{equation}

\begin{equation}
M = D H^{1/\delta}, \ for\ T = T_c
\end{equation}

where $\epsilon = (T-T_c)/T_c$; $M_0$, $h_o/m_0$, D are critical amplitudes; and $\beta,\gamma, \delta$ are critical exponents\cite{Fisher67}. For the original Arrott plot, $\beta$ = 0.5 and $\gamma$ = 1\cite{Arrott}. In a more general case with different critical exponents, a modified Arrott plot is often considered, which formulates as:\cite{Arrott67}. 

\begin{equation}
(\frac{H}{M})^{1/\gamma} = a\epsilon + b M^{1/\beta}
\end{equation}

where a and b are fitting constants. Since the mean field values apparently do not agree with our experimental data, we adopt the modified Arrott plot in order to better understand the nature of this ferromagnetic transition. In Fig.~\ref{AT1}(b), we thus show the modified Arrott plot with a set of parameters ($\gamma$ = 1.38, $\beta$ = 0.25) that produces a set of straight lines satisfying the Arrott criteria. The ferromagnetic transition temperature, as depicted by the data that passes through the origin, is $T_c=$71 K.

\begin{figure}[tbh!]
\includegraphics[scale = 0.6]{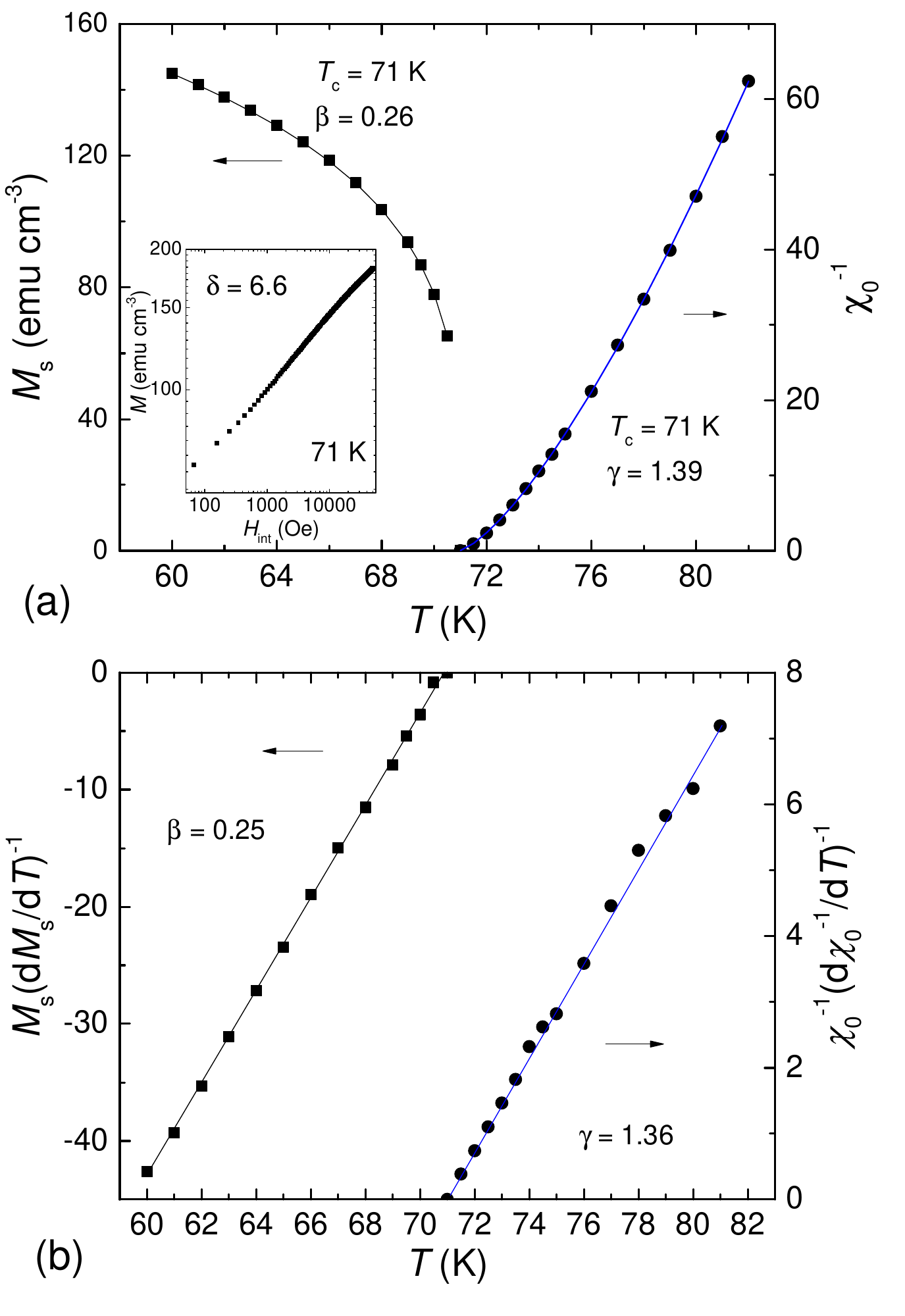}
\caption{(Color online) Temperature dependent spontaneous magnetization $M_s$ and inverse initial magnetic susceptiblity $\chi_0^{-1}$. Inset shows the field-dependent magnetization data measured at $T_c$ = 71 K on a log-log plot. (b) Kouvel-Fisher plots for $M_s(dM_s/dT)^{-1}$ and $\chi_0^{-1}(d\chi_0^{-1}/dT)^{-1}$. Solid lines are fitting curves to the data.}
\label{AT2}
\end{figure}

In order to double-check the self-consistency of obtained critical exponents\cite{Banerjee09}, we extract the spontaneous magnetization and initial magnetic susceptibility from Fig.~\ref{AT1}(b). The linearly extrapolated $M_s$ and $\chi^{-1}_0$ are plotted as a function of temperature in Fig.~\ref{AT2}(a). The solid curves are fitted lines according to Equation 1 and 2. The fitted critical exponent values are $\gamma$ = 1.39 and $\beta$ = 0.26, which are similar to the starting values. In the inset of Fig.~\ref{AT2}(a), the field-dependent magnetization of CrSbSe$_3$ at $T_c$ = 71 K is drawn on a log-log plot. According to Equation 3, the field dependence yield $\delta$ = 6.6. In comparison to the theoretical prediction based on the Widom relation\cite{Widom64}:

\begin{equation}
\delta = 1 + \frac{\gamma}{\beta}
\end{equation}

the calculated value of $\delta$, using experimentally obtained critical exponents $\gamma$ = 1.38 and $\beta$ = 0.25, is $\sim$6.5, which agrees with what is obtained in Fig.~\ref{AT2}(a). 

The self-consistency can also be checked via the Kouvel-Fisher (KF) method\cite{KF64}. According to the KF method,

\begin{equation}
\frac{M_s(T)}{dM_s(T)/dT} = \frac{T-T_c}{\beta}
\end{equation}

\begin{equation}
\frac{\chi_0^{-1}(T)}{d\chi_0^{-1}(T)/dT} = \frac{T-T_c}{\gamma}
\end{equation}

both $\frac{M_s(T)}{dM_s(T)/dT}$ and $\frac{\chi_0^{-1}(T)}{d\chi_0^{-1}(T)/dT}$ are linear function of temperature, with their slopes equal to the inverse of critical exponents. As seen in Fig.~\ref{AT2}(b), linear lines are fitted to the experimental data and yield $\gamma$ = 1.36, $\beta$ = 0.25, again, very similar to the starting critical exponent values.

The self-consistency of these critical exponents indicates a good fit of the data with theoretical formalism. The set of critical exponents obtained is not close to any of the well-known theoretical values for various models. The values are also quantitatively different from that obtained for layered ternary chromium tri-chalcogenides\cite{Lin17, Liu17}. The use of the modified Arrott plot for fitting critical exponents is known to yield significant standard errors in some cases, which arises when choosing different field and temperature ranges for consideration. As a consequence, the standard errors are usually difficult to estimate, as has been discussed recently, for example, in CrGeTe$_3$\cite{Lin17, Liu17}, where slightly different criteria result in different critical exponent values. For the current study on CrSbSe$_3$, it is clear that the mean field critical exponents cannot describe the experimental data. The obtained critical exponents serve as a set of self-consistent values that matches our data, and may direct further detailed studies to a range around these values. 

Finally, according to the Mermin-Wagner theorem, ferromagnetic or antiferromagnetic ordering cannot happen in ideal one- or two-dimensional systems at finite temperature within an isotropic Heisenberg model\cite{Mermin66}. In the case where magnetic anisotropy exists, however, this conclusion does not necessarily hold. In the case of CrSbSe$_3$, despite the fact that the magnetic-bearing sublattice of Cr$^{3+}$ appears to be quasi-one-dimensional, the observed magnetic anisotropy in the ordered ferromagnetic state may contribute to its stability. Nevertheless, the obtained critical exponents still likely reflect the influence of low-dimensionality of the Cr$^{3+}$ sublattice; further, more detailed investigation may be of interest. 

\section{Conclusions}

In summary, we have synthesized CrSbSe$_3$ single crystals in a high-temperature, Se-rich solution and carried out a study on the structural, electrical and magnetic properties of that material. Structurally, CrSbSe$_3$ shows a pseudo-one-dimensional structure with magnetic-moment-bearing Cr sublattice forming double rutile chains. Electronically, CrSbSe$_3$ is a semiconductor with a band gap of $\sim$0.7 eV. In its paramagnetic state, it appears to be magnetically isotropic with an effective moment of $\sim$4.1 $\mu_B$/Cr and a Curie-Weiss temperature of $\sim$145 K. The ferromagnetic transition temperature was determined from a modified Arrott plot as $T_c$ = 71 K. In the ferromagnetic state, CrSbSe$_3$ is anisotropic with the $a$-axis being the easy axis. Based on the modified Arrott plot and the Kouvel-Fisher method, we arrived at a set of critical exponents for second order phase transition that describe our experimental data: $\gamma$ = 1.38, $\beta$ = 0.25 and $\delta$ = 6.6. These values could serve as a starting point for further theoretical studies on magnetism at low-dimensions.

\section*{Acknowledgements}
We would like to thank Valentin Taufour for insightful discussions and Tia S. Lee for experimental assistance. This work was supported by the Gordon and Betty Moore EPiQS initiative, grant number GBMF-4412.

\section*{Appendix}

The SXRD data was collected at 296 K with a Kappa APEX DUO diffractometer equipped with a CCD detector (Bruker) using graphite-monochromatized Mo-K$_\alpha$ radiation ($\lambda$ = 0.71073 $\text{\AA}$). The raw data were corrected for background, polarization, and the Lorentz factor using APEX2 software\cite{Bruker2013}, and a multi-scan absorption correction was applied\cite{Sheldrick14}. The structure was solved using the charge flipping method\cite{Suto04} and subsequent difference Fourier analyses with Jana2006\cite{Oszlanyi08,Petricek14,petvrivcek2014b}. Structure refinement against F$_o^2$ was performed with Shelxl-2017/1\cite{Sheldrick15,Sheldrick17}.

\begin{table}[h!]
\caption{Crystallographic data and details of the structure determination of CrSbSe$_3$ derived from single-crystal experiments measured at 296(1) K.}
\begin{tabular}{p{4cm} p{4.5cm}}
\hline
Sum Formula & CrSbSe$_3$\\
\hline
Formula weight/(g mol$^{-1}$) & 410.63\\
Crystal system & orthorhombic\\
Space group & $Pnma$ (no. 62)\\
Formula units per cell, $Z$ & 4\\
Lattice parameter a/ $\text{\AA}$ & 9.1388(3)\\
$b$/ $\text{\AA}$ & 3.7836(1)\\
$c$/ $\text{\AA}$ & 13.3155(4)\\
Cell volume/ ($\text{\AA}^3$) & 463.88(2)\\
Radiation & $\lambda$(Mo-$K_{\alpha}$ = 0.71073 $\text{\AA}$)\\
& 2$\theta \leq$ 82.23$^{\circ}$\\
Data range & -16$\leq$h$\leq$16, -6$\leq$k$\leq$6, -23$\leq$l$\leq$23\\
Absorption coefficient/mm$^{-1}$ & 31.5\\
Measured reflections & 30213\\
Independent reflections & 1527\\
Reflections with $I$ $>$ 2$\sigma$($I$) & 1379\\
$R$(int)& 0.040\\
$R$($\sigma$) & 0.015\\
No. of parameters & 31\\
$R_1$(obs) & 0.017\\
$R_1$(all $F_0$) & 0.022\\
$wR_2$(all $F_0$) & 0.030\\
Residual electron density (e$^-$/$\text{\AA}^3$) & 1.12 to -0.98\\
\hline
\end{tabular}
\label{table1}
\end{table}

\begin{table}[h!]
\caption{Wyckoff positions, coordinates, occupancies, and equivalent displacement parameters respectively for CrSbSe$_3$ single-crystal measured at 296(1) K. $U_{eq}$ is one third of the trace of the orthogonalized $U_{ij}$ tensor}
\centering
\begin{tabular}{p{1.0cm} p{1.0cm} p{1.5cm} p{0.7cm} p{1.5cm} p{0.8cm} p{1.5cm}}
\hline
Atom & Wyck. site & $x$ & $y$ & $z$ & Occ. & $U_{eq}$\\
Sb1 & 4$c$ & 0.47058(2) & 1/4 & 0/65793(2) & 1 & 0.01332(3)\\
Cr1 & 4$c$ & 0.15547(4) & 3/4 & 0.54468(3) & 1 & 0.00701(5)\\
Se1 & 4$c$ & 0.28484(3) & 3/4 & 0.71314(2) & 1 & 0.01009(4)\\
Se2 & 4$c$ & 0.00186(2) & 3/4 & 0.39112(2) & 1 & 0.00784(4)\\
Se3 & 4$c$ & 0.32801(2) & 1/4 & 0.48446(2) & 1 & 0.00918(4)\\
\hline

\end{tabular}
\label{table2}
\end{table}

\begin{table}[h!]
\caption{Anisotropic displacement parameters for CrSbSe$_3$ single-crystal measured at 296(1) K. The coefficients $U_{ij}$ (/$\text{\AA}^2$) of the tensor of the anisotropic temperature factor of atoms are defined by exp[–2$\pi^2$($U_{11}$h$^2$a$^{*2}$ + $\dots$ + 2$U_{23}$klb$^*$c$^*$)]}
\centering
\begin{tabular}{p{1.0cm} p{1.80cm} p{1.8cm} p{1.8cm} p{1.8cm}}
\hline
Atom & $U_{11}$ & $U_{22}$ & $U_{33}$ & $U_{13}$\\
\hline
Sb1 & 0.01182(6) & 0.00899(6) & 0.01915(7) & -0.00246(5)\\
Cr1 & 0.00735(12) & 0.00616(12) & 0.00752(13) & -0.00026(10)\\
Se1 & 0.01217(9) & 0.00866(8) & 0.00946(9) & -0.00064(7)\\
Se2 & 0.00848(8) & 0.00777(8) & 0.00726(8) & 0.00029(7)\\
Se3 & 0.00937(8) & 0.00815(8) & 0.01001(9) & 0.00127(7)\\
\hline

\end{tabular}
\label{table3}
\end{table}

\bibliographystyle{apsrev4-1}
%

\end{document}